\documentclass[runningheads]{llncs}
\usepackage{graphicx}

\usepackage{comment}
\usepackage{amsmath,amssymb} 

\usepackage[accsupp]{axessibility}  
\usepackage{graphicx}
\usepackage{amsmath,amssymb}
\usepackage{multirow}
\usepackage[table]{xcolor}

\usepackage{listings}
\usepackage{subcaption}
\usepackage{algorithm}
\usepackage{algpseudocode}

\usepackage{txfonts}
\usepackage[T1]{fontenc}

\begin{document}
\pagestyle{headings}
\mainmatter
\def\ECCVSubNumber{069}  

\title{FUSION: Fully Unsupervised Test-Time Stain Adaptation via Fused Normalization Statistics} 


\titlerunning{FUSION}
%


\author{Nilanjan Chattopadhyay$^*$ \and
Shiv Gehlot$^*$  \and
Nitin Singhal}
%
\authorrunning{Chattopadhyay et al.}
%
\institute{Advanced Technology Group, AIRA MATRIX, India \\ \email{\{nilanjan.chattopadhyay,shiv.gehlot,nitin.singhal\}@airamatrix.com} 
}
\maketitle
\def\thefootnote{*}\footnotetext{These authors contributed equally}
\begin{abstract}
Staining reveals the micro-structure of the aspirate while creating the histopathology slides. Stain variation, defined as a chromatic difference between the source and the target, is caused by varying characteristics during staining, resulting in a distribution shift and poor performance on the target. The goal of stain normalization is to match the target's chromatic distribution to that of the source. However, stain normalisation causes the underlying morphology to distort, resulting in an incorrect diagnosis. We propose FUSION, a new method for promoting stain-adaption by adjusting the model to the target in an unsupervised test-time scenario, eliminating the necessity for significant labelling at the target end. FUSION works by altering the target's batch-normalization statistics and fusing them with source statistics using a weighting factor. The algorithm reduces to one of two extremes based on the weighting factor. Despite the lack of training or supervision, FUSION surpasses existing equivalent algorithms for classification and dense predictions (segmentation), as demonstrated by comprehensive experiments on two public datasets.
\end{abstract}

\section{Introduction}
Staining is used in histopathology slide preparation to highlight the essential structure of the aspirate. Differences in stain chemicals, lighting conditions, or staining time may produce stain colour variances in images collected at the same or separate facilities. The inter-center chromatic difference results in sub-optimal performance on the target test set because of stain variation, limiting model deployment across the centres.

Transfer learning is a simple but effective strategy for dealing with distribution shifts, but it requires annotation efforts at the target end, which can be challenging, especially in medical image analysis. By training a network to learn stain variance via mapping color-augmented input images to the original images, self-supervised learning eliminates the need for annotation \cite{Tellez2019,shivssl}. However, the settings of the train-time augmentations influence the performance of these algorithms on the target domain. \emph{Stain normalisation} is a stain variation handling method that does not require any training at the source or target end. It aligns the source and target chromatic distributions using reference image(s) from the source domain \cite{kothari2011automatic,mccann2014algorithm,ruifrok2001quantification,abe2005color,GUPTA2020,macenko2009method,reinhard2001color,ruderman1998statistics,magee2009colour}. However, depending on the reference image(s), its performance can vary greatly. Changes in the underlying structure of the images further degrade the performance. The use of generative modeling-based solutions eliminates the need for a reference image. However, it requires the data from target domain in addition to a larger training sample, which limits its application \cite{IsbiZanjani2018,MidlZanjani2018,stainGan}. 

These limitations are overcome via domain adaptation \cite{Ganin2015UnsupervisedDA,TzengHDS15} and test time training \cite{TTT}, but with a modified training process. The goal of adaptation is to generalise a model $f^s_{\theta}(x)$ trained on source data $\{x^s,y^s\}$ to target data $x^t$. Test time adaptation (TTA) modifies the testing procedure while preserving the initial training. Two TTA techniques explored in the literature are Entropy Minimization (EM) \cite{wang2021tent,memo2021} and Normalization Statistics Adaptation (SNA) \cite{vanillabna,snb}. In EM, the backpropagation to update the adaptation parameters $\theta$ is driven by the entropy function formulated using the predictions of $f^s_{\theta}(x)$ on $x^t$. Batch normalisation statistics and parameters are controlled in \cite{wang2021tent} by entropy minimization of the target's prediction. For gradient update, \cite{memo2021} minimises the entropy of the marginal distribution computed from several augmented versions of a single test sample. SNA eliminates the need for backpropagation by focusing on batch normalization-specific statistics (mean and variance) and only requires forward steps. During the inference in \cite{vanillabna}, the batch normalisation statistics are updated with $x^t$. On the other hand, \cite{snb} updates statistics with $x^t$ considering $x^s$ as a prior.

FUSION, the proposed methodology, performs batch statistics adaptation as well. FUSION is a generic technique that combines source and target batch normalisation layers via a weighting factor that favors the source or the target end. The stain fluctuation between $x^s$ and $x^t$ is likewise connected to the weighting factor. FUSION outperforms the conventional SNA techniques because of its generic nature. The domain difference between $x^s$ and $x^t$ is represented in this study by stain variation; thus, the terms ``\emph{model adaptation}'' and ``\textit{stain adaptation}'' are used interchangeable.

\if
Stain variation in histological images is ubiquitous with the most common being hematoxylin and eosin (H$\&$E). Such variation occurs due to variety of reasons such as differences in chemicals, the time and amount of dye applied to the tissue on the glass, section thickness, and scanner. Although standard staining protocols are often employed, in practice, the variation in image contrast and color among samples persists. Figure 1(a) illustrates the stain variation present among the samples of breast lymph nodes from different institutions.

State-of-the-art deep learning models are vulnerable to these stain variation. Models when trained with data from one center fail to generalize to images from other centers with different stains. This is one of the greatest obstacle in widespread adaptation of neural networks in the histology labs and clinics. Many advancements have been made in the deep learning field to combat this sub-optimal generalization performance. These approaches can be majorly grouped into the following two categories: stain normalization during inference which can utilize label free data of different stain or data augmentation during training such that model can learn domain-invariant features.
\fi

\begin{figure}[!t]
    \centering
    \centerline{
    \subfloat[$x^s$ and $x^t$]{
    \includegraphics[height=2.6 cm, width=2.6 cm]{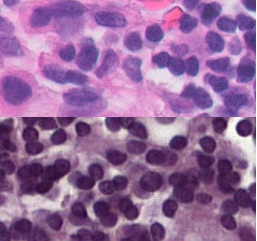}
    \label{mugshot_c4}}
    \subfloat[Vanilla Inference]{
    \includegraphics[height=2.6 cm, width= 3 cm]{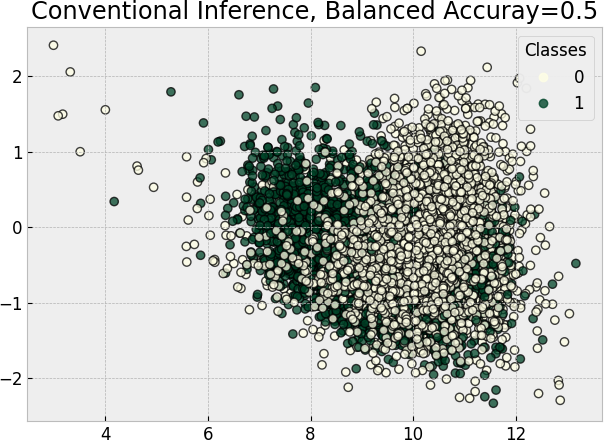}}
    \subfloat[FUSION ($\beta=0.6$)]{
    \includegraphics[height=2.6 cm, width= 3.2 cm]{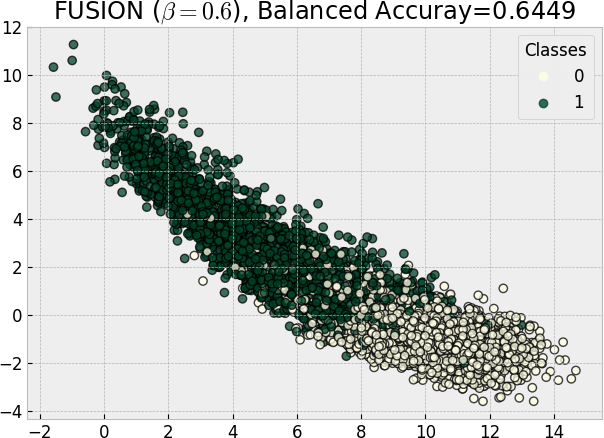}}
    \subfloat[FUSION ($\beta=0.9$)]{
    \includegraphics[height=2.6 cm, width= 3.2 cm]{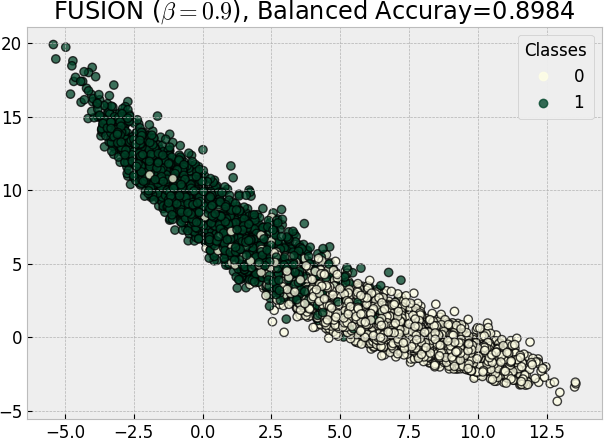}}
    }
    \caption{The vanilla inference performs poorly on the target due to considerable stain variation between $x^s$ (source) and $x^t$ (target). FUSION may bypass this constraint by combining the $x^s$ and $x^t$ normalisation statistics, resulting in significant performance gains. }
    \label{paper_mugshot}
\end{figure}
\if
Data augmentation is a proven way to improve the robustness of computer vision models. They try to boost the robustness of model in two ways - 1. increase diversity of the training data, 2. increase the hardness of the training data. But there are few shortcomings when relying solely on data augmentation to improve model robustness to unseen stain variation. Light data augmentation doesn't introduce enough diversity to improve performance whereas aggressive data augmentation can increase the hardness to such an extent it starts hurting performance even on training set. Fine tuning the hyper-parameters to find the optimal mix of diversity and hardness is cumbersome and not practical. We are most likely to use sub-optimal data augmentation policies and hyper-parameters that is not tailored to a unseen stains. Though this improves the robustness of the model to a great extent, it is often not enough to stop the model from performance deterioration.

Stain normalization is another widely used methodology for improving model performance for histological images with different stains. It works by trying to change the color distribution of the test data(target data) to that of the training images(source data). After changing the unknown stain to a known stain, we expect to see a improvement in performance since the model is now processing WSIs with training stain. Stain normalization methodology can be used while training as data augmentation combined with other augmentations, or only while doing inference on images with different stain. Though in principle, this methodology can close the gap between train performance vs performance on target data, in practice it can fall short due to variety of reasons. One such reason is that stain normalization often requires manual selection of a representative image from the source data to which the color distribution of the entire target data is matched. This not ideal since a single image can't capture the variability present in training data and can deliver inaccurate normalization results. Another major issue is introduction of artifacts after stain conversion. The artifacts results in loss of performance countering the gain that could have been achieved by stain normalization. 

To combat such issues, we propose a new methodology where instead of changing the image's chromatic distribution, the coviarate shift caused by the changed stain is corrected internally and automatically using BatchNorm. This novel methodology of model adaption results in improved robustness to out-of-distribution data in a fully-unsupervised test-time setting without introducing any artifacts in the input images. In Table 1 we compare our methodology with other model adaption methods as well as methods that are commonly used to deal with stain variation like stain normalization and color augmentations. We demonstrate significant gains on data from other centers with different stains outperforming other methodologies. We also show how stain change results in loss of performance and correlate the magnitude of the performance drop loss with activation distribution changes inside a deep learning model.
\fi

\if
Adaptation aims to generalize a model $f^s_{\theta}(x)$ trained using the source data $\{x^s,y^s\}$ on the target data $x^t$. Transfer learning is a naive yet effective approach to target adaptation but demands annotation efforts at the target end, limiting its applicability. Domain adaptation and test time training discard this limitation, albeit with a modified training procedure. Test time adaptation (TTA) specifically alters the testing procedure, holding the initial training intact. Entropy Minimization (EM) and  Normalization Statistics Adaptation (SNA) are two discussed TTA approaches in the literature. In EM, the entropy function formulated using the predictions  of $f^s_{\theta}(x)$ on $x^t$ drives the backpropagation to update the adaptation parameters $\theta$. SNA nullifies backpropagation requirements by targeting batch normalization-specific statistics (mean and variance), requiring only the forward steps.  If the deviation between $x^s$ and $x^t$ exists in terms of stain variation, adaptation can be interpreted as \textit{stain adaptation}, which is the focus of this work. \fi

\section{FUSION}
Batch normalisation (BN) is built into current Convolutional Neural Networks (CNNs) for stable and faster training. The second-order batch statistics ($\mu$, $\sigma$) and two learnable parameters for scaling and shifting are used by BN layers to normalise the features of each batch.
\begin{equation}
    \hat{g}(x_B^s)=\gamma^s \frac{g(x_B^s)-\mathbb{E}[g(x_B^s)]}{\sqrt{(Var[g(x_B^s)]+\epsilon}} + \alpha^s,
    \label{batch_norm_train}
\end{equation}
where, $x_B^s$ is a training batch from $x^s$, and $g(x_B^s)$ are activations. The moving averages of $\mu$ and $\sigma$ are also kept during the training phase to use during inference.

These statistics can differ greatly between two centres (source and target) in the case of stain variation (Fig.~\ref{paper_mugshot}). As a result, in an \emph{inter-center training-testing} system, the training statistics may lead to poor performance \cite{vanillabna} during inference on test set from a different center. 
However, we believe that the batch's ($x_B$) requirement to represent the target data distribution $p_{test}(x)$ is important for classification because a batch may not contain samples from all classes. Accordingly, for inference on $x^t$, Eq. \ref{batch_norm_train} is modified as:
\begin{equation}
    \hat{g}(x_B^t)=\gamma^s \frac{g(x_B^t)-\mathbb{E}[g(x_B^t)]}{\sqrt{(Var[g(x_B^t)]+\epsilon}} + \alpha^s; \qquad \qquad x_B^t \in p_{test}(x)
    \label{batch_norm_test}
\end{equation}
Even though each test batch represents the test data distribution, it is not sufficient to describe the statistics of the complete test set.
Although all of the test samples can be pooled into a single batch, this has computational limits. We describe a computationally efficient strategy for leveraging the statistics of the complete test set during inference. The Eq. \ref{batch_norm_test} is utilized for inference in the \textit{first step}, and running estimates of $\mu$ and $\sigma$ produced from the test-batches are kept at the same time (similar to training). In the \textit{second step}, the collected running averages are utilized in the final inference, altering Eq.~\ref{batch_norm_test} to:
\begin{equation}
    \hat{g}(x_B^t)=\gamma^s \frac{g(x_B^t)-\mathbb{M}^t}{\sqrt{(\mathbb{V}^t+\epsilon}} + \alpha^s,
    \label{batch_norm_final}
\end{equation}
where, $\mathbb{M}^t$ and $\mathbb{V}^t$ represents the running statistics of mean and variance, respectively in the \textit{first step}. 
\begin{figure}[!t]
    \centering
    \centerline{
    \includegraphics[scale=.77]{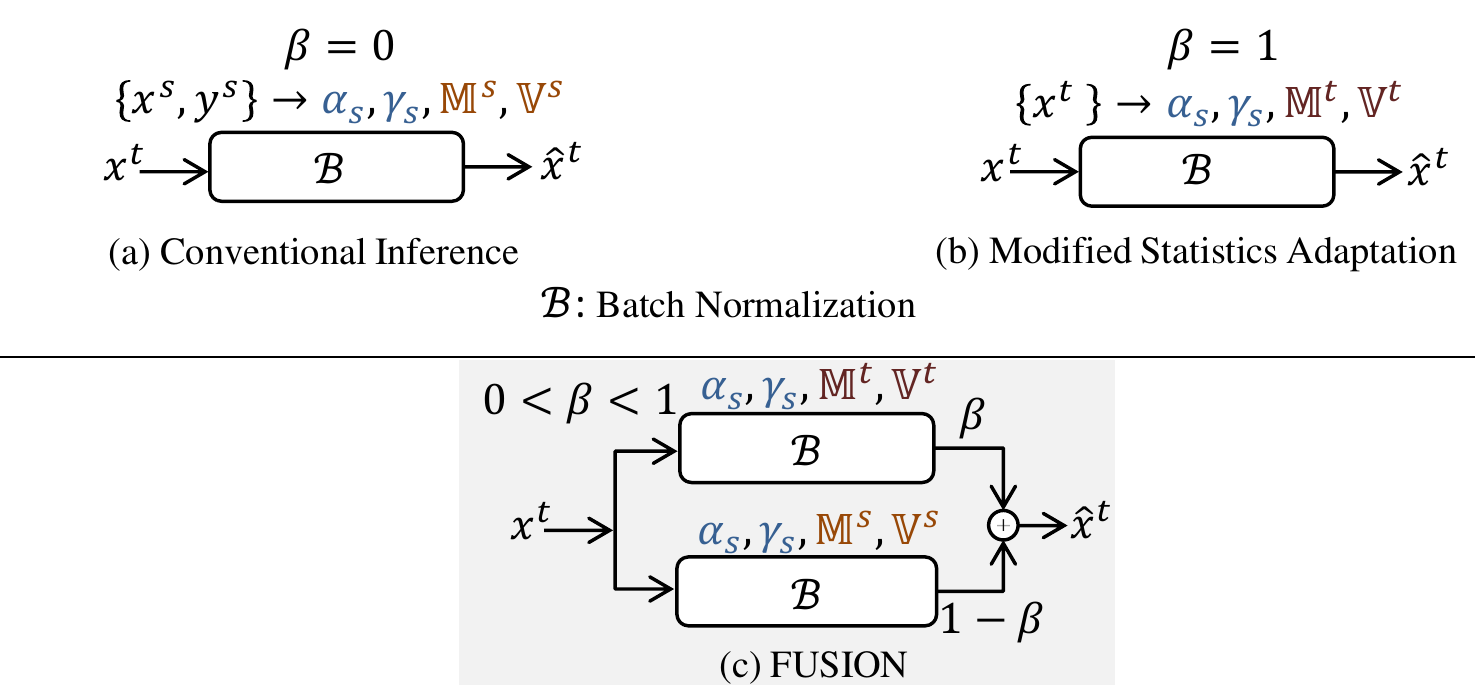}}
    \caption{Conventional inference (a) uses the running statistics of the source $(\mathbb{M}^s, \mathbb{V}^s)$ for inference on target $x^t$. The modified statistics adaptation approach (b) replaces it with the target statistics, $\mathbb{M}^t, \mathbb{V}^t$. Unlike prior approaches to statistics adaptation, which only considered per-batch statistics, this method considers the complete target set. FUSION (c) is a generalized statistics adaptation approach that fuses the source and target statistics with a weighting factor $\beta$, reducing to (a) and (b) for $\beta=0$ and $1$, respectively.}
    \label{method_fig_1}
\end{figure}
\paragraph{\textnormal{\textbf{Fusing the Batch Normalizations}}} Only the statistics of the target are considered in Eq. \ref{batch_norm_final}, completely discarding  the statistics of the source. Excessive perturbations of the source statistics may cause the performance to deteriorate. As a result, optimal performance can be achieved by combining both statistics. To this purpose, Eq. \ref{batch_norm_final} is updated as follows:

\begin{equation}
    \hat{g}(x_B^t)=\gamma^s \left(\beta \left[ \frac{g(x_B^t)-\mathbb{M}^t}{\sqrt{(\mathbb{V}^t+\epsilon}}\right]+(1-\beta) \left[ \frac{g(x_B^t)-\mathbb{M}^s}{\sqrt{(\mathbb{V}^s+\epsilon}}\right]\right) + \alpha^s,
    \label{batch_norm_fused}
\end{equation}
$\beta \in [0,1]$ is a hyperparameter, $\mathbb{M}^s$ and $\mathbb{V}^s$ represents moving averages for source. Eq. \ref{batch_norm_fused} is a generalized test time statistic normalization adaptation. For $\beta=1$, it reduces to Eq. \ref{batch_norm_final}, considering only the target statistics. While for $\beta=0$, it focuses only on the source statistics. With $0<\beta<1$ as the weighting parameter, Eq. \ref{batch_norm_fused} exploits both the domains.


\section{Experiments}
\begin{figure}[!t]
    \centering
    \centerline{
    \subfloat[C0]{
    \includegraphics[scale=0.15]{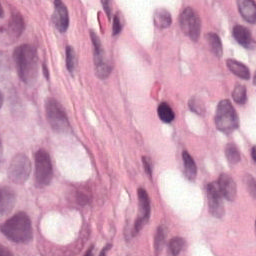}}
    \subfloat[C1]{
    \includegraphics[scale=0.15]{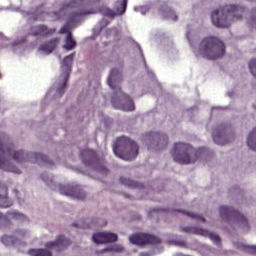}}
    \subfloat[C2]{
    \includegraphics[scale=0.15]{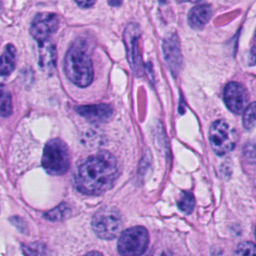}}
    \subfloat[C3]{
    \includegraphics[scale=0.15]{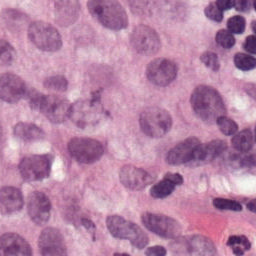}}
    \subfloat[C4]{
    \includegraphics[scale=0.15]{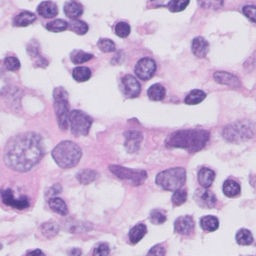}}
    \qquad
    \subfloat[C0]{
    \includegraphics[scale=0.30]{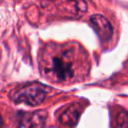}}
     \subfloat[C1]{
    \includegraphics[scale=0.30]{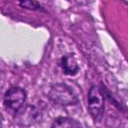}}
     \subfloat[C2]{
    \includegraphics[scale=0.30]{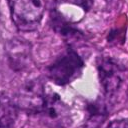}}
    }
    \caption{Sample images from different centers of Camelyon17 (a-e) and TUPAC (f-h), highlighting the inter-center stain variation. }
    \label{sample-images}
\end{figure}

\textbf{Applications:} FUSION is evaluated on \emph{classification} and \emph{segmentation} tasks using Camelyon -17 and TUPAC datasets. Classification and segmentation are performed on the five-center Camelyon-17 \cite{CAMELYON17} dataset, but only classification is performed on the three-center TUPAC dataset. Sample images and a detailed dataset description may be found in Table \ref{dataset-description} and Fig \ref{sample-images}, respectively. Datasets vary widely among locations, making it probable that a model developed for one location may perform badly when applied to other locations. The effectiveness of FUSION in preventing such performance decline is put to the test.\\
\textbf{Implementation:} ResNet-18 \cite{resnet} and EfficientNet-B0 \cite{efnet} are trained for classification with and without Train-time augmentations (TrTAug). Rotation-based augmentation is the default, whereas TrTAug also includes color-based augmentations such as hue, saturation, and value (HSV) variations to induce stain invariance. The Feature Pyramid Network (FPN) \cite{fpn} with ResNet-34 \cite{efnet} as an encoder is utilized for the segmentation task. Each network is trained with SGDM optimizer for 55 epochs and a step LR scheduler of 0.1 decay factor. The learning rate was set to .001, the batch size to 64, and the weight decay to .01. For FUSION, optimal $\beta$ is selected through grid search from $[0.6, 0.7, 0.8, 0.9]$.\\
\textbf{Baselines:} We exhibit the influence of test-time augmentations (TTAug), stain normalisation (\cite{macenko2009method,vahadane}), and normalisation statistics adaption (\cite{vanillabna,snb}) in addition to vanilla inference. The \cite{vanillabna} and \cite{snb} are represented as vanilla batch normalization adaptation (Vanilla-BNA) and sample based BNA (SB-BNA), respectively. The influence of Eq. \ref{batch_norm_test} in conjunction with Vanilla-BNA (Vanilla-BNA + Eq. \ref{batch_norm_test}) is also investigated. FUSION-full or FUSION with $\beta=1$ (Eq. \ref{batch_norm_final}) is also analyzed. N is set to 20 for SB-BNA, as suggested in \cite{snb}.

\begin{table}[!t]
\caption{Multi centers (sources) datasets are used for classification and segmentation to analyze inter-center stain adaptation with FUSION. For each dataset, highlighted center is considered as source and remaining as the targets. }
\label{dataset-description}
\centerline{
\resizebox{\textwidth}{!}{
\begin{tabular}{c|cccccc|c}
\hline
Application                      & \multicolumn{6}{c|}{Dataset (Centers)}                                                                                                                                                                                   & Size                   \\ \hline
                                 & \multicolumn{1}{c|}{}                              & \multicolumn{1}{c|}{C0}    & \multicolumn{1}{c|}{C1}                         & \multicolumn{1}{c|}{\cellcolor[HTML]{C0C0C0}C2} & \multicolumn{1}{c|}{C3}    & C4    &                        \\ \cline{3-7}
                                 & \multicolumn{1}{c|}{\multirow{-2}{*}{Camelyon-17 \cite{CAMELYON17}}} & \multicolumn{1}{c|}{13527} & \multicolumn{1}{c|}{9690}                       & \multicolumn{1}{c|}{14867}                      & \multicolumn{1}{c|}{30517} & 26967 & \multirow{-2}{*}{256 $\times$ 256}  \\ \cline{2-8} 
                                 & \multicolumn{1}{c|}{}                              & \multicolumn{2}{c|}{\cellcolor[HTML]{C0C0C0}C0}                              & \multicolumn{2}{c|}{C1}                                                      & C2    &                        \\ \cline{3-7}
\multirow{-4}{*}{Classification} & \multicolumn{1}{c|}{\multirow{-2}{*}{TUPAC \cite{tupac}}}       & \multicolumn{2}{c|}{4260}                                                    & \multicolumn{2}{c|}{1600}                                                    & 1344  & \multirow{-2}{*}{128 $\times$ 128}  \\ \hline
                                 & \multicolumn{1}{c|}{}                              & \multicolumn{1}{c|}{C0}    & \multicolumn{1}{c|}{\cellcolor[HTML]{C0C0C0}C1} & \multicolumn{1}{c|}{C2}                         & \multicolumn{1}{c|}{C3}    & C4    &                        \\ \cline{3-7}
\multirow{-2}{*}{Segmentation}   & \multicolumn{1}{c|}{\multirow{-2}{*}{Camelyon-17 \cite{CAMELYON17} }} & \multicolumn{1}{c|}{237}   & \multicolumn{1}{c|}{231}                        & \multicolumn{1}{c|}{300}                        & \multicolumn{1}{c|}{437}   & 685   & \multirow{-2}{*}{1024 $\times$ 1024} \\ \hline
\end{tabular}}}
\end{table}

\section{Results}
\begin{figure}[!t]
    \centering
    \centerline{
    \subfloat[]{
    \includegraphics[scale=0.47]{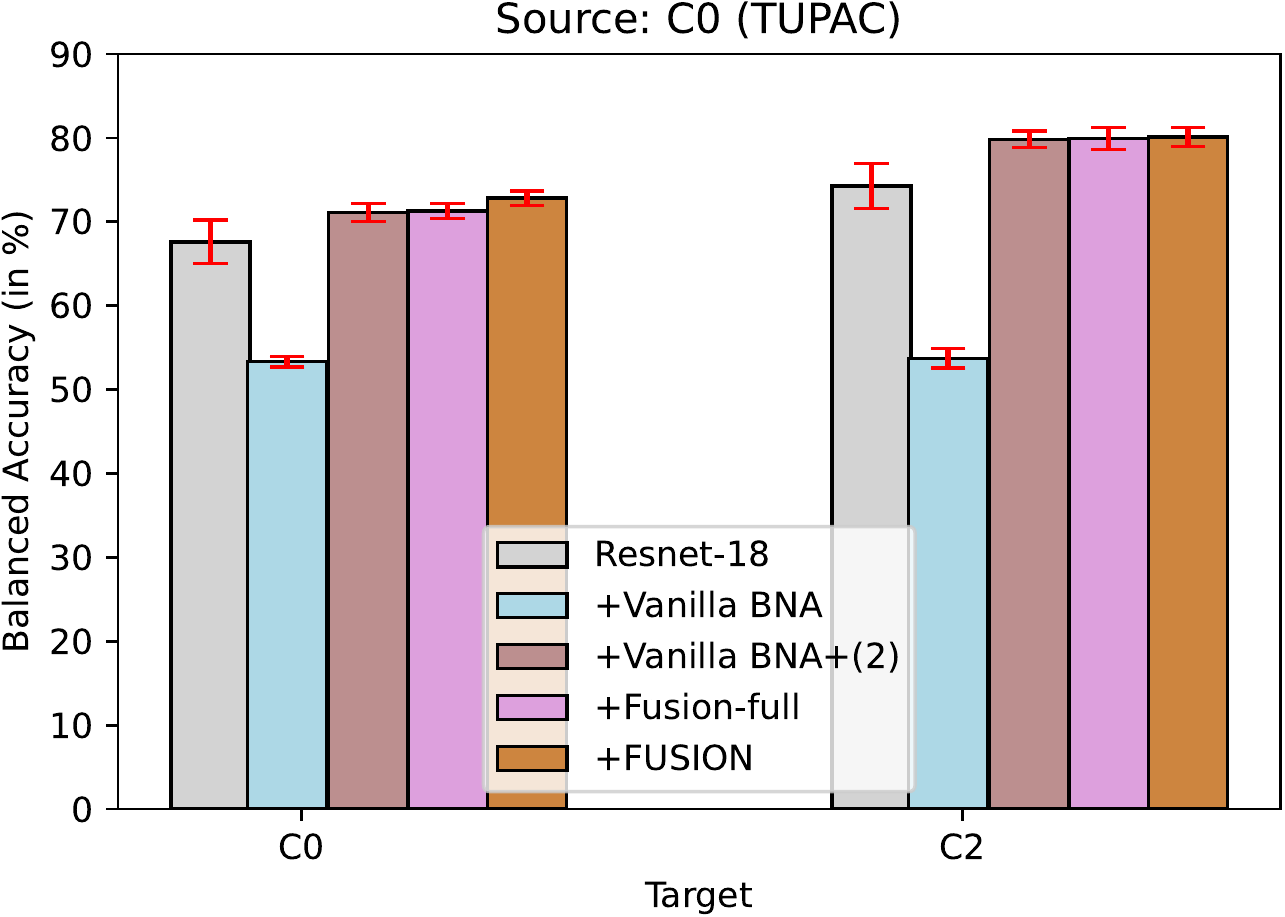}}
    \subfloat[]{
    \includegraphics[scale=0.47]{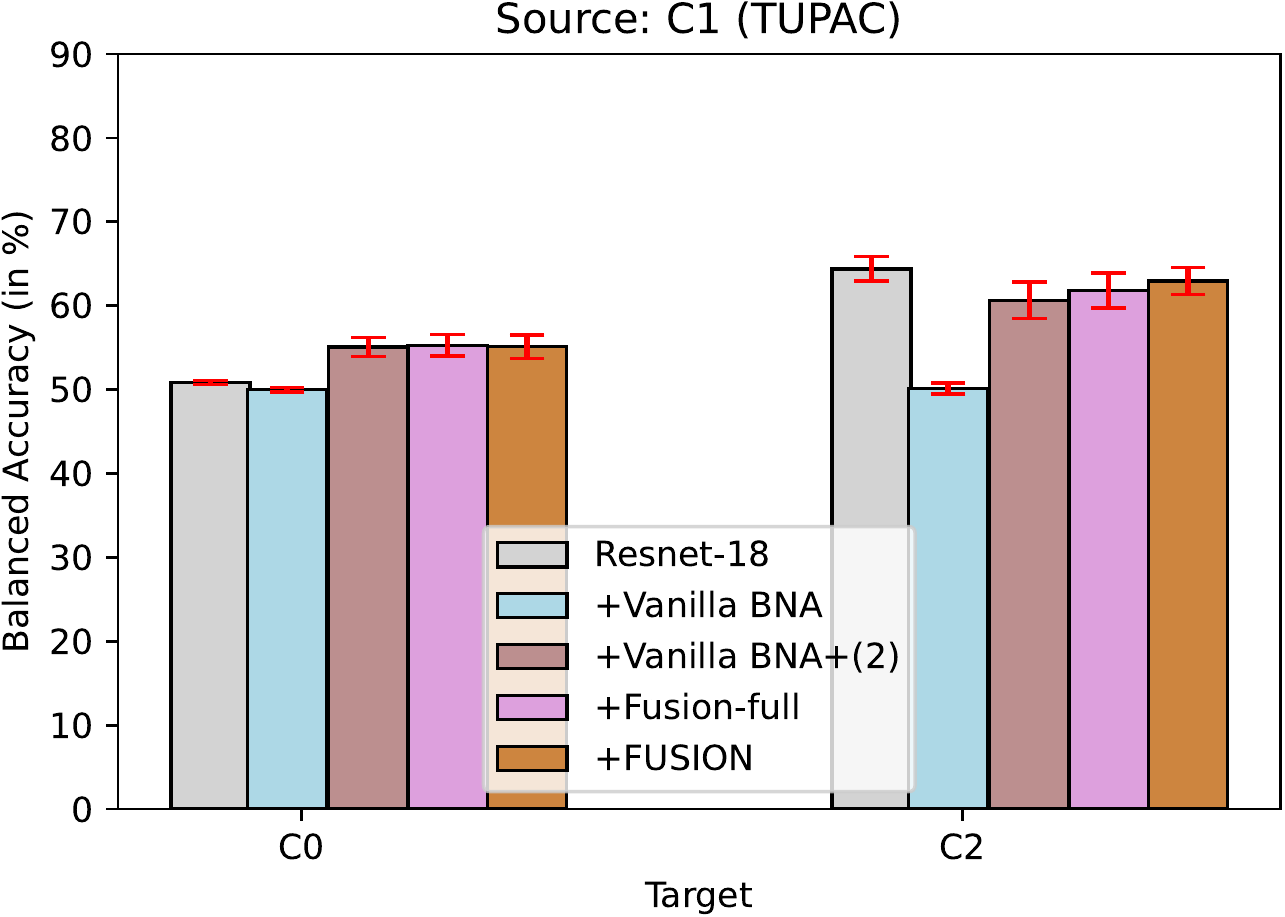}}
    }
    \caption{\emph{Balanced accuracy} for \emph{classification} on TUPAC \cite{tupac} with source C0 (a) and C1 (b). Vanilla BNA, like Camelyon-17 in Table \ref{classification-results}, performs worse than vanilla inference in the absence of Eq. \ref{batch_norm_test}. In certain circumstances, FUSION-full and FUSION are comparable to Vanilla BNA, while in some cases, they are superior. Also, with C1 as the source and C2 as the target, FUSION produces the least decrement, proving the validity of statistics merging.}
    \label{tupac-results}
\end{figure}

\paragraph{\textnormal{\textbf{FUSION gives the best classification and segmentation results in varied test setups.}}} Table \ref{classification-results} indicates that conventional inference results in poor performance. This is attributed to drastic stain variation between C2 (source) and other centers (targets), implying the necessity of stain adaptation. FUSION outperforms the other approaches when it comes to classification on Camelyon-17 under stain-variation condition. The performance improvement for C0, C1, C3, and C4 in terms of balanced accuracy for EfficientNet-B0 (without TrTAug) is 43.26 percent, 22.89 percent, 38.85 percent, and 27.60 percent, respectively. Introducing TrTAug, as expected, introduces stain invariance in the network, resulting in increased performance across target centers. FUSION outperforms other methods and provides highest increment over the baseline with TrTAug. Similarly, FUSION is the optimal technique for TUPAC dataset, as shown in Fig. \ref{tupac-results}. It performs on par to FUSION-full in a few cases while superior in others.

FUSION is the best performing approach for segmentation (Table \ref{segmentation-results}), with the highest gain of 8.55 percent in terms of dice score for C0. Performance on C4 shows decrease as seen in classification as well. However, FUSION results in least decrement in comparison to other methods. As can be seen, FUSION-full underperforms FUSION in most cases, implying that full adaptation is inefficient and that an optimal combination of $x^s$ and $x^t$ statistics is required. 

\setlength{\tabcolsep}{4pt}

\begin{table}[!t]
\caption{\textit{Balanced accuracy} for \textit{classification} on Camelyon-17 \cite{CAMELYON17} with  C2 as the source. Best results are highlighted in bold. Maximum increment and minimum decrement are represented in blue and red, respectively. Vanilla BNA has poor performance without Eq. \ref{batch_norm_test}. Hence, for classification, each batch at inference time must represents the test set ($x_B^t \in p_{test}(x)$).}
\label{classification-results}
\resizebox{\textwidth}{!}{%
\begin{tabular}{lcccc}
\hline
\multicolumn{1}{c}{Source} & \multicolumn{4}{c}{Target}                                                    \\ \hline
\multicolumn{1}{c}{C2}     & C0                               & C1           & C3           & C4           \\ \hline
EfficientNet-B0 \cite{efnet}           & {50.34 $\pm$ 0.57} & 49.43 $\pm$ 0.64 & 49.90 $\pm$ 0.13 & 63.99 $\pm$ 2.67 \\
\hspace{1 em}+ TTAug             & 51.07 $\pm$ 1.27 ($\uparrow 0.73$)                & 43.37 $\pm$ 1.08 ($\downarrow 6.06$) & 44.73 $\pm$ 0.94 ($\downarrow 5.17$) & 54.06 $\pm$ 0.71 ($\downarrow 9.93$) \\
\hspace{1 em}+ Macenko \cite{macenko2009method}            & 50.21 $\pm$ 0.19 ($\downarrow 0.13$)                & 50.32 $\pm$ 2.84 ($\uparrow 0.89$) & 49.83 $\pm$ 3.03 ($\downarrow 0.07$) & 49.95 $\pm$ 0.05 ($\downarrow 14.04$) \\
\hspace{1 em}+ Vahadane \cite{vahadane} & 84.65 $\pm$ 2.20 ($\uparrow 30.31$) & 76.29 $\pm$ 3.26 ($\uparrow 26.86$) & 76.67 $\pm$ 3.13 ($\uparrow$ 26.77) & 46.08 $\pm$ 4.35 ($\downarrow 17.91$) \\
\hspace{1 em}+ Vanilla BNA \cite{vanillabna}              & 53.47 $\pm$ 0.19    ($\uparrow$ 3.13)                & {53.27 $\pm$ 0.24} ({$\uparrow$ 3.84}) & 53.47 $\pm$ 0.24 ($\uparrow$ 3.57) & 51.55 $\pm$ 0.17 ($\downarrow$ 12.44) \\
\hspace{1 em}+ Vanilla BNA \cite{vanillabna} +(\ref{batch_norm_test})              & 92.43 $\pm$ 0.24 ($\uparrow 42.09$)                & 72.11 $\pm$ 0.40 ($\uparrow 
22.69$) & 87.29 $\pm$ 0.52 ($\uparrow 37.39$) & 91.44 $\pm$ 0.73 ($\uparrow 27.45$) \\
\hspace{1 em}+ SB-BNA \cite{snb}            & 67.94 $\pm$ 5.33
 ($\uparrow 17.6$)                & 62.32 $\pm$ 2.35 ($\uparrow 12.98$) & 57.52 $\pm$ 2.57 ($\uparrow 7.62$) & 85.67 $\pm$ 1.46 ($\uparrow 21.68$) \\
\hspace{1 em}+ FUSION-full  (\ref{batch_norm_final})                 & 92.61 $\pm$ 0.12 ($\uparrow 42.27$)                   & 71.27 $\pm$ 0.37 ($\uparrow 21.84$) & 87.59 $\pm$ 0.31 ($\uparrow 37.69$) & 92.14 $\pm$ 0.76 ($\uparrow 28.15$) \\
\hspace{1 em}+ FUSION (\ref{batch_norm_fused})                & \textbf{93.59 $\pm$ 0.13} (\textcolor{blue}{$\uparrow$ 43.25})                 & \textbf{72.32 $\pm$ 0.54} (\textcolor{blue}{$\uparrow 22.89$}) & \textbf{88.75 $\pm$ 0.32} (\textcolor{blue}{$\uparrow 38.85$}) & \textbf{91.59 $\pm$ 0.90} (\textcolor{blue}{$\uparrow 27.60$})\\

\rowcolor[HTML]{F0E7E6} 
+ TrTAug              & 93.32 $\pm$ 1.11                     & 75.28 $\pm$ 3.66 & 85.28 $\pm$ 1.51 & 82.01 $\pm$ 3.12 \\
\rowcolor[HTML]{F0E7E6} 
\hspace{1 em}+ TTAug             & 92.25 $\pm$ 0.99 ($\downarrow 1.07$)                & 75.03 $\pm$ 2.66
 ($\downarrow 0.25$) & 82.57 $\pm$ 0.88 ($\downarrow 2.71$) & 79.02 $\pm$ 2.42 ($\downarrow 2.99$) \\ \rowcolor[HTML]{F0E7E6} 
\hspace{1 em}+ Macenko  \cite{macenko2009method}           & 92.72 $\pm$ 1.12 ($\downarrow 0.60$)                & 82.98 $\pm$ 3.09 ($\uparrow 7.70$) & 83.80 $\pm$ 3.24 ($\downarrow 1.48$) & 83.75 $\pm$ 1.83 ($\uparrow 1.74$) \\
\rowcolor[HTML]{F0E7E6} 
\hspace{1 em}+ Vahadane \cite{vahadane} & 93.41 $\pm$ 0.78 ($\uparrow$ 0.09) & 80.53 $\pm$ 1.73 ($\uparrow$ 5.25) & 88.79 $\pm$ 0.59 ($\uparrow$ 3.21) & 71.37 $\pm$ 1.18 ($\downarrow$ 10.64) \\
\rowcolor[HTML]{F0E7E6} 
\hspace{1 em}+ Vanilla BNA \cite{vanillabna}            & 53.78 $\pm$ 0.18 ($\downarrow 39.54$)                & 53.03 $\pm$ 0.23 ($\downarrow 22.25$) & 53.75 $\pm$ 0.13 ($\downarrow 31.53$) & 51.86 $\pm$ 0.20 ($\downarrow 30.15$) \\ \rowcolor[HTML]{F0E7E6} 
\hspace{1 em}+ Vanilla BNA \cite{vanillabna} +(\ref{batch_norm_test})               & 94.29 $\pm$ 0.16 ($\uparrow$ 0.97)                  & 83.41 $\pm$ 1.08 ($\uparrow$ 8.13) & 91.67 $\pm$ 0.22 ($\uparrow$ 6.39) & 94.81 $\pm$ 0.22 ($\uparrow$ 12.8) \\
\rowcolor[HTML]{F0E7E6} 
\hspace{1 em}+ SB-BNA \cite{snb}              & 93.96 $\pm$ 0.79 ($\uparrow 0.64$)                  & 83.87 $\pm$ 1.01 ($\uparrow 8.59$) & 90.45 $\pm$ 0.74 ($\uparrow 5.17$) & 93.22 $\pm$ 0.72 ($\uparrow 11.21$) \\
\rowcolor[HTML]{F0E7E6}
\hspace{1 em}+ FUSION-full  (\ref{batch_norm_final})                 & 94.69 $\pm$ 0.22   ($\uparrow 1.37$)                  & 83.83 $\pm$ 1.22 ($\uparrow 8.55$) & 91.93 $\pm$ 0.29 ($\uparrow 6.65$) & 95.38 $\pm$ 0.18 ($\uparrow 13.38$) \\
\rowcolor[HTML]{F0E7E6} 
\hspace{1 em}+ FUSION   (\ref{batch_norm_fused}).               & \textbf{95.28 $\pm$ 0.33} (\textcolor{blue}{$\uparrow 1.96$})                   & \textbf{88.68 $\pm$ 0.87} (\textcolor{blue}{$\uparrow$ 13.4}) & \textbf{93.02 $\pm$ 0.34} (\textcolor{blue}{$\uparrow$ 7.74})& \textbf{95.58 $\pm$ 0.12} (\textcolor{blue}{$\uparrow$ 13.57})\\ \hline

\end{tabular}
}
\end{table}

\setlength{\tabcolsep}{4pt}

\begin{table}[!t]
\caption{\textit{Dice Score} for \textit{segmentation} on Camelyon-17 \cite{CAMELYON17} with  C1 as the source. FUSION has maximum increment for C0, C2, C3, and least decrement for C4. Due to dense predictions, Eq. \ref{batch_norm_test} is inherently satisfied, making its combination with Vanilla BNA redundant.}
\label{segmentation-results}
\resizebox{\textwidth}{!}{%
\begin{tabular}{lcccc}
\hline
\multicolumn{1}{c}{Source} & \multicolumn{4}{c}{Target}                                                    \\ \hline
\multicolumn{1}{c}{C1}     & C0                               & C2           & C3           & C4           \\ \hline
ResNet-34 \cite{resnet}           & {62.75 $\pm$ 7.95} & 53.49 $\pm$ 14.00 & 58.23 $\pm$ 8.87 & 50.52 $\pm$ 12.61 \\
\hspace{1 em}+ TTA             & 56.23 $\pm$ 6.65 ($\downarrow 6.52$)                & 47.30 $\pm$ 8.87 ($\downarrow 6.19$) & 50.46 $\pm$ 7.01 ($\downarrow 7.77$) & 39.60 $\pm$ 8.91 ($\downarrow 10.92$) \\
\hspace{1 em}+ Vanilla BNA \cite{vanillabna}             & 70.56 $\pm$ 4.36    ($\uparrow$ 7.81)                & {55.70 $\pm$ 3.11} ({$\uparrow$ 2.21}) & 59.80 $\pm$ 3.44 ($\uparrow$ 1.57) & 38.05$\pm$ 4.80 ($\downarrow$ 10.92) \\
\hspace{1 em}+ Vanilla BNA \cite{vanillabna}+(\ref{batch_norm_test})              & 70.47 $\pm$ 4.23 ($\uparrow 7.72$)                & 56.08 $\pm$ 3.57 ($\uparrow 2.59$) & 59.85 $\pm$ 3.57 ($\uparrow 1.62$) & 38.26 $\pm$ 4.59 ($\downarrow 12.26$) \\
\hspace{1 em}+ FUSION-full (\ref{batch_norm_final})                 & 70.36 $\pm$ 5.00 ($\uparrow 7.61$)                   & 57.08 $\pm$ 6.17 ($\uparrow 3.59$) & 60.62 $\pm$ 3.79 ($\uparrow 2.39$) & 40.95 $\pm$ 6.73 ($\downarrow 9.57$) \\
\hspace{1 em}+ FUSION (\ref{batch_norm_fused})                & \textbf{71.30 $\pm$ 5.95} (\textcolor{blue}{$\uparrow$ 8.55})                 & \textbf{60.68 $\pm$ 9.66} (\textcolor{blue}{$\uparrow 7.19$}) & \textbf{63.97 $\pm$ 4.81} (\textcolor{blue}{$\uparrow 5.74$}) & \textbf{48.06 $\pm$ 7.40} (\textcolor{red}{$\downarrow 2.46$})\\ \hline
\end{tabular}
}
\end{table}
\paragraph{\textnormal{\textbf{In Vanilla BNA, each batch must be representative of the test data distribution.}}} In the absence of Eq. \ref{batch_norm_test}, vanilla BNA performs worse than vanilla inference. Batch normalisation statistics update without Eq.~\ref{batch_norm_test} does not examine the impact of all classes in the dataset, resulting in poor performance. As pixels in a single image belong to many classes, this is satisfied by default for segmentation. Also, because Eq. \ref{batch_norm_final} considers updates due to all test samples (statistical running averages) throughout inference, Eq. \ref{batch_norm_test} is not required separately.


\paragraph{\textnormal{\textbf{FUSION restores activation distribution shifts caused by stain differences.}}}

BatchNorm normalizes the distribution of activations over a minibatch during training. By setting the first two moments (mean and variance) of the distribution of each activation to be zero and one, it tries to correct for the changes in distribution of activations of a layer in the network caused by update of parameters of the previous layers. This distributional changes in layer inputs after every minibatch is known as internal covariate shift.

By default, a running mean and variance of activations for all minibatches is tracked in BatchNorm layer to be used during test time such that the activation mean is 0 and variance is 1. But when the model is used for prediction on an unseen and different stain, the distribution of activations changes and behaves differently from training time. The existing BatchNorm statistics, since based on different distribution, can't correct for the new distribution shift. This causes the new mean and variance of activations after BatchNorm layer differ from the intended values of mean 0 and variance 1.

Let $\mu_c^l$ be the running mean and $\sigma_c^l$ be the variance recorded during training at layer $l$ for channel $c$. With stain variation between $x^s$ and $x^t$: 
\begin{align}
    \mathbb{E}[g_c^l(x^s)] & \ne \mathbb{E}[g_c^l({x^t})] \\
    \mathsf{Var}[g_c^l(x^s)] & \ne \mathsf{Var}[g_c^l({x^t})]
    \label{different_distribution_var}
\end{align}
After applying BatchNorm the above equations can be written as:
\begin{align}
    \mathbb{E}[\mathsf{BN}(g_c^l(x^s))] & \ne \mathbb{E}[\mathsf{BN}(g_c^l({x^t}))] \\
    \mathsf{Var}[\mathsf{BN}(g_c^l(x^s))] & \ne \mathsf{Var}[\mathsf{BN}(g_c^l({x^t}))]
    \label{same_distribution}
\end{align}
FUSION is designed to restore these activations in a way that:
\begin{align}
    \mathbb{E}[\mathsf{BN}(g_c^l(x^s))] & \approx \mathbb{E}[\mathsf{FUSION}(g_c^l({x^t}))] \\
    \mathsf{Var}[\mathsf{BN}(g_c^l(x^s))] & \approx \mathsf{Var}[\mathsf{FUSION}(g_c^l({x^t}))]
    \label{same_distribution}
\end{align}
where FUSION can be calculated using Eq. \ref{batch_norm_fused}.

\if
This is because when images with same stain is passed through model during inference, activations at layer $l$ for channel $c$ over the minibatches holds the following equations:
\begin{equation}
    \mathbb{E}[\frac{g_c^l(x)-\mu_c^l}{\sigma_c^l}] \approx 0
    \label{running_mean}
\end{equation}
\begin{equation}
    \mathsf{Var}[\frac{g_c^l(x)-\mu_c^l}{\sigma_c^l}] \approx 1
    \label{running_var}
\end{equation}
But the image with different stain is passed:
\begin{equation}
    \mathbb{E}[\frac{g_c^l(x)-\mu_c^l}{\sigma_c^l}] \approx 0
    \label{running_mean}
\end{equation}
\begin{equation}
    \mathsf{Var}[\frac{g_c^l(x)-\mu_c^l}{\sigma_c^l}] \approx 1
    \label{running_var}
\end{equation}
\fi

\begin{figure}[!t]
\centering
\centerline{
\subfloat[Channel 325 \label{Covariate-Shift1}]{
\includegraphics[height= 4 cm, width= 4 cm]{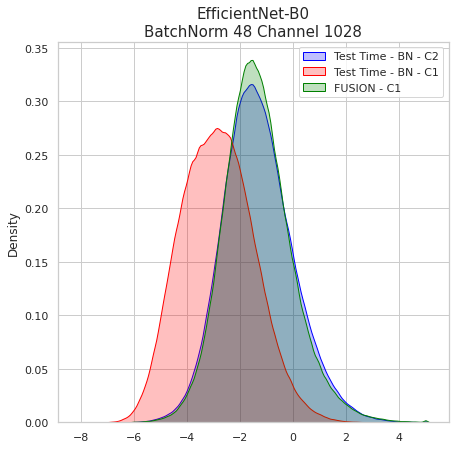}}
\subfloat[Channel 1028 \label{Covariate-Shift2}]{
\includegraphics[height= 4 cm, width= 4 cm]{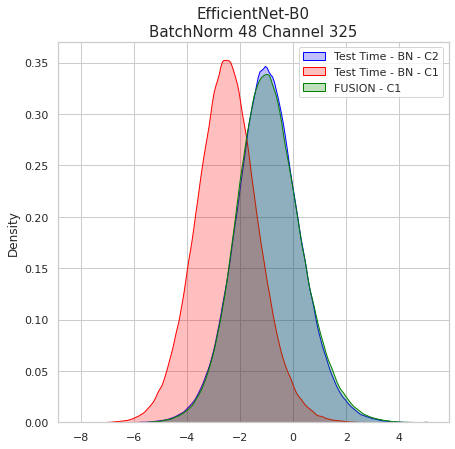}}
\subfloat[Channel 892 \label{Covariate-Shift3}]{
\includegraphics[height= 4 cm, width= 4 cm]{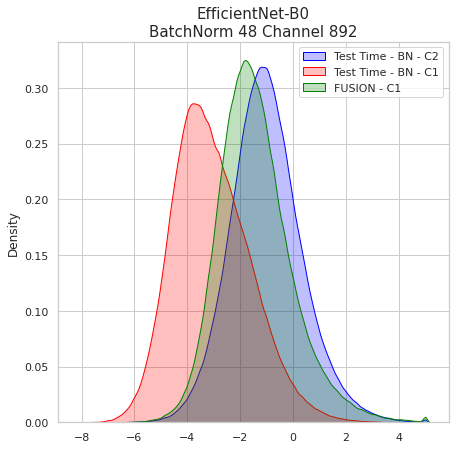}}
}
\caption{(a-c) Due to a stain difference between the source and target, the feature map distribution shifts with EfficientNet-B0 after the BatchNorm layer. After the last BatchNorm layer, the activations are recovered from the EfficientNet-B0 model. C2 was used to train the model at the beginning. The blue line displays the density plot of a channel using source validation data, while the red line represents the shift owing to the target's different stain. FUSION aims to match the training distribution by correcting the covariate shift (green line).}
\end{figure}

The shift in activation distribution (Fig. \ref{Covariate-Shift1}, \ref{Covariate-Shift2}, \ref{Covariate-Shift3}) that explains the performance decrease with stain variation is correlated to the stain difference. With similar stains, the distribution shift is minor, and source batch normalisation statistics are sufficient. With substantial stain variation, however, the shift is significant with major reduction in test performance, and FUSION corrects it back to the source. As a result, FUSION's achieves higher performance improvement even on greater stain difference (higher distribution shift).


\begin{figure}[!t]
\centering
\centerline{\subfloat[Impact of step count \label{BatchAnalysis-1}]{
\includegraphics[height= 5 cm, width= 7 cm]{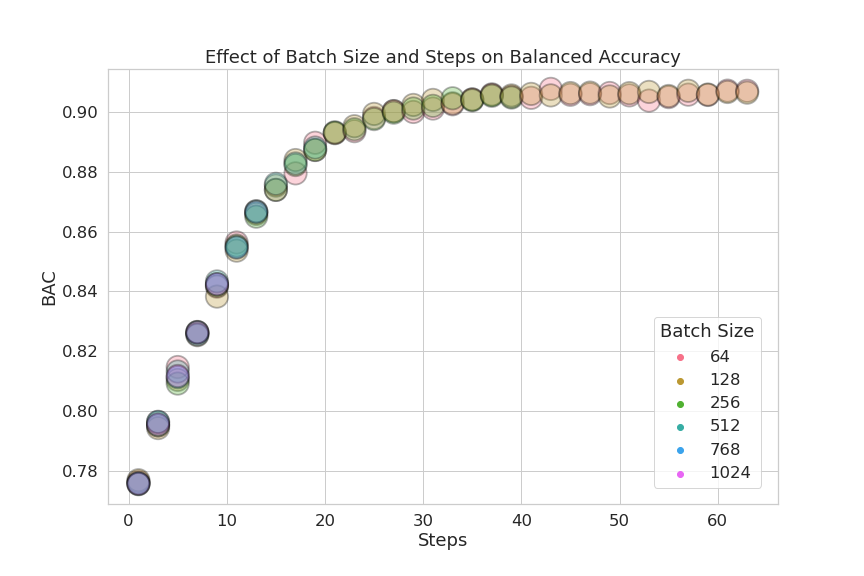}}
\subfloat[Impact of batch size \label{BatchAnalysis-2}]{
\includegraphics[height= 5 cm, width= 4.5 cm]{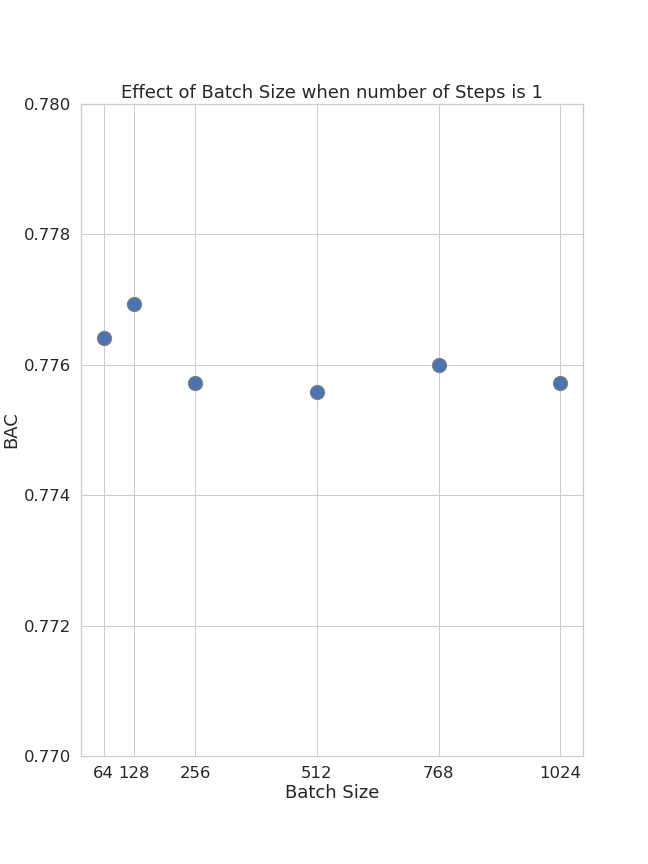}}}
\caption{Impact of (a) steps count and (b) batch size on FUSION's performance in terms of balanced accuracy. The model is trained on \emph{C2} with varying batch size and steps count, and mean balanced accuracy is reported on remaining test centers (centers \emph{C0, C1, C3, C4}). 
(a) FUSION's performance is related to BatchNorm statistics which correlates to batch size and number of iterations/steps. Increased steps counts improves the performance irrespective of batch sizes. As moving average of mean and variance is used for updating BatchNorm statistics, large number of steps are required to overcome the affect of momentum. 
(b) When applying FUSION, batch size has negligible effect on performance when step count is fixed (set to 1). The difference in performance is not dependent on batch size. The similar observations can be made in (a) where points of different batch sizes overlap each other for same step count.}
\end{figure}

\paragraph{\textnormal{\textbf{{The performance of FUSION correlates to the number of steps.}}}}

Batch size and step counts have an impact on batch normalisation statistic calculations. The performance of updating statistics from the target does not improve as batch size is increased. The amount of steps taken to generate the new statistics, on the other hand, is critical. As shown in Fig. \ref{BatchAnalysis-1}, the performance of different batch sizes converges roughly in the same way as step count increases. This could be related to the reduced influence of momentum while using step counts for batch statistics upgradation. Further evidence of importance of step count vs batch size can be seen in Fig. \ref{BatchAnalysis-2} where for the same step count, batch size has no correlation with performance.

\paragraph{\textnormal{\textbf{{FUSION is generic and provides flexibility during inference.}}}}  Full adaptation is neither generic nor efficient, as it may result in substantial statistical deviation. Similarly, poor performance without adaptation is due to incompatibility of the source statistics with the target. FUSION's generalizability, which allows it to focus on any source or target via the weighting factor $\beta$, is a key advantage. FUSION provides more flexibility during inference because of its entirely unsupervised and test-time adaptive nature.

\section{Conclusion} FUSION was proposed for entirely unsupervised test-time stain adaptation, and its performance was evaluated in various scenarios. Due to its unique qualities that combine source and target batch normalisation statistics, FUSION outperforms similar approaches in the literature. The rigorous testing with various datasets, applications, and architectures reveals that FUSION provides optimal performance. The characteristics of FUSION, such as its impact on countering the covariate shift, also justify the obtained higher performance. Analysis of batch size and the step counts impact on FUSION during inference highlights its dependency on step counts instead of batch size. However, as suspected, batch size of one is not sufficient for FUSION to work as batch normalization statistics calculation is required. Apart from its performance, FUSION provides inference time flexibility to steer it towards the source or target through a single hyperparameter. Hence, it allows no adaptation, full adaptation, or partial adaptation.

\clearpage
%
%
\bibliographystyle{splncs04}
\bibliography{069}
\end{document}